\documentclass{aa}
\usepackage{graphicx}                    
\usepackage{color}                       
\usepackage{breakurl}                         

\newcommand{\proba}{PROBA2}
\newcommand{\lyra}{LYRA}
\newcommand{\xrs}{GOES/XRS}
\newcommand{\goes}{GOES}
\newcommand{\eit}{SOHO/EIT}
\newcommand{\trace}{TRACE}

\newcommand{\xmin}{$x_{min}$}
\newcommand{\specialcell}[2][l]{\begin{tabular}[#1]{@{}l@{}}#2\end{tabular}}

\begin{document}

\title{Effects of flare definitions on the statistics of derived flare distributions}

\author{D.\ F.~Ryan\inst{1,2} \and M.~Dominique\inst{1} \and 
            D.~Seaton\inst{1,3,4} \and K.~Stegen\inst{1} \and A.~White\inst{5}
           }

\institute{Solar-Terrestrial Center of Excellence, SIDC, Royal Observatory of Belgium, Brussels, Belgium \and NASA Goddard Space Flight Center, Greenbelt, Maryland, USA \and Cooperative Institute for Research in Environmental Sciences, University of Colorado, Boulder, Colorado, USA \and NOAA National Centers for Environmental Information, Boulder, Colorado, USA \and School of Computer Science and Statistics, Trinity College Dublin, O'Reilly Institute, Dublin 2, Ireland}

\abstract{
The statistical examination of solar flares is crucial to revealing their global characteristics and behaviour.  Such examinations can tackle large-scale science questions or give context to detailed single-event studies.  However, they are often performed using standard but basic flare detection algorithms relying on arbitrary thresholds.  This arbitrariness may lead to important scientific conclusions being drawn from results caused by subjective choices in algorithms rather than the true nature of the Sun.  In this paper, we explore the effect of the arbitrary thresholds used in the \goes\ (Geostationary Operational Environmental Satellite) event list and \lyra\ (Large Yield RAdiometer) Flare Finder algorithms.  We find that there is a small but significant relationship between the power law exponent of the \goes\ flare peak flux frequency distribution and the  flare start thresholds of the algorithms.  We also find that the power law exponents of these distributions are not stable, but appear to steepen with increasing peak flux.  This implies that the observed flare size distribution may not be a power law at all.  We show that depending on the true value of the exponent of the flare size distribution, this deviation from a power law may be due to flares missed by the flare detection algorithms.  However, it is not possible determine the true exponent from \xrs\ observations.  Additionally we find that the \proba/\lyra\ flare size distributions are artificially steep and clearly non-power law.  We show that this is consistent with an insufficient degradation correction.
This means that \proba/\lyra\ should not be used for flare statistics or energetics unless degradation is adequately accounted for.  However, it can be used to study variations over shorter timescales and for space weather monitoring.
}
\keywords{Sun: corona, Sun: flares, Sun: X-rays, gamma rays, Methods: statistical, Methods: data analysis}

\maketitle

\section{Introduction}
\label{sec:intro}
Solar flares are amongst the most energetic events in the solar system, releasing up to 10$^{32}$~erg in minutes or hours.  They are believed to arise from a rapid re-ordering of highly stressed magnetic fields in the corona \citep{shib2011}, but this process is still not fully understood.  One way of better understanding the flaring process and the role it plays in the corona is to examine flares in a statistical way.

For example, the flaring process may hold the key to the coronal heating problem: the observation that the lower corona is 2--3 orders of magnitude hotter than the photosphere ($\sim$2~MK as opposed to $\sim$6,000~K).  It is thought that ubiquitous small-scale unresolved flaring events, known as nanoflares, may liberate enough energy from the stressed coronal magnetic field to maintain the observed high coronal temperatures.  Although this is not the only proposed solution -- others involve the dissipation of magnetohydrodynamic (MHD) waves -- much work has been done in both modelling the required nanoflare distribution and analysing the observed flare distribution \citep[see reviews by][]{klim2006,real2010,parn2012}.  One method has been to assume that the flare energy size distribution follows a power law which holds over all scales.  The power law exponent can then be determined from larger observed flares and extrapolated to the nanoflare regime.  \citet{huds1991} calculated that the critical value of the power law exponent is 2.  If the exponent is greater than this then there is enough energy in the nanoflare ensemble to sustain the corona's temperature.  If the exponent is less than this value, then some other mechanism is required.

Another reason to explore the statistical flare distribution is to better understand the flaring process itself.  One theory used to try to explain how flares occur is self-organised criticality (SOC).  The first SOC models of solar flares were devised by \citet{lu1991} and \citet{lu1993} and draw from the earlier theoretical work of \citet{bak1987,bak1988}.  Since then numerous variations have been devised \citep[see review by][]{char2001}.  These models hold that solar flares are in fact an avalanche of small-scale magnetic energy release events.  This avalanche can be triggered once the magnetic fields in the corona -- which slowly grow in complexity owing to the jostling motions of the photospheric plasma in which they are rooted -- surpass some critical instability threshold.  One of the key predictions of these models is that the flare size or frequency distribution is scale invariant and characterised by a power law \citep{asch2016}.  Although SOC is not the only process that can lead to power law size distributions -- others include self-organisation, inverse-cascades, stochastic relaxation, turbulence\ -- a power law size distribution nonetheless remains a fundamental aspect of SOC models of solar flares.  Therefore, fully understanding the functional form of the flare size distribution is crucial to testing these models and better understanding the flaring process.

There have been many previous studies which have examined the flare size distribution in soft X-rays \citep[e.g.\ ][]{huds1969, drak1971, shimi1995, lee1995, feld1997, shimo1999, asch2000, vero2002a, vero2002b, yash2006, asch2012}.  Many previous studies have examined the flare peak size distribution rather than the total energy size distribution.  There are a number of reasons for this.  First, no single instrument can adequately determine the energy of the entire flare.  Second, the duration of some flares are not fully observed because of the instrument duty cycle or because temporally overlapping flares are very difficult to separate in spatially integrated observations.  However, if it is true that the flare peak flux scales with total energy, then the peak can be used as a proxy.  Under this assumption, the studies mentioned above have found power law slopes in the range 1.64--2.16 with most studies finding a slope less than 2, but not necessarily beyond statistical uncertainty.  Other studies \citep[e.g.][]{berg1998, parn2000, asch2000, benz2002} have attempted to more directly observe the small-scale brightenings in the quiet Sun corona that may be responsible for coronal heating.  These studies used extreme ultraviolet (EUV) imaging observations from the Transition Region And Coronal Explorer (\trace) and/or the Extreme ultraviolet Imaging Telescope on board the Solar and Heliospheric Observatory (\eit).  The power law indices of the energy size distributions of the observed events vary considerably depending on the study.  \citet{berg1998} and \citet{asch2000} found power law slopes or 1.35 and 1.8, implying that coronal heating by nanoflares is highly unlikely.  However, \citet{parn2000} and \citet{benz2002} found power law slopes of 2.52 and 2.31, which strongly support the coronal heating by nanoflares picture.  The cause of these discrepancies remains contested.  \citet{benz2002} attributed the shallower slopes of other studies to  biases in their energy calculation and event selection criteria.  In contrast, \citet{asch2002} claimed that the steeper slopes were due to temperature biases of the image passbands used.  Thus, the importance of the flare distribution for coronal heating has  not yet been conclusively answered.

To date, one aspect of these studies has rarely been discussed in detail: how does the way flares are defined affect the  statistics of the derived size distribution?  Solar flare event lists are typically generated by simple detection algorithms relying on arbitrary thresholds.  The choice of these thresholds could affect not only the power law exponent of the derived flare size distribution, but also whether it is a power law in the first place.  Moreover many studies do not provide statistical evidence that their data are truly represented by a power law.  Although some studies do statistically test their data \citep[e.g.\ ][]{whea2010}, others only provide a visual observation of a straight-lined log-log histogram.  However, it has been shown that first justifying and then fitting a power law model to data via graphical methods can be misleading  \citep{clau2009,dhuy2016}.  In light of the importance of the power law nature of the flare distribution to flare theories (e.g.\ those based on self-organised criticality) and the value of the power law exponent (e.g.\ for coronal heating), it is crucial to properly understand the effect of flare detection thresholds and the nature of the derived flare size distribution before drawing any physical conclusions.

An example of a study which has examined the effect of event definition on the statistics of derived distributions is \citet{buch2005}.  They constructed a number of possible types of event definitions inspired by methods used in the literature.  They applied these different definitions to time series of energy dissipation from an MHD turbulence simulation.  They then examined the size distributions of various properties of the detected events (total energy, peak energy, duration, etc.) and found that their derived distributions changed significantly depending on how the events were defined.

In this paper, we ask a similar question, but apply it to real flare observations and real detection algorithms.  We examine the effect of the arbitrary thresholds used in the \goes\ event list and \lyra\ Flare Finder (LYRAFF) flare detection algorithms when applied to time-series of \xrs\ (Geostationary Operational Environmental Satellite/X-Ray Sensor) and \proba/\lyra\ (PRoject for OnBoard Autonomy 2/Large Yield RAdiometer) observations.  The algorithms are run on the same data sets  using different threshold values.  We then perform rigorous statistical analyses on the the resulting flare peak size distributions in order to determine how sensitive they are to the choice of algorithm thresholds.  The results lead us to investigate the power law nature of such distributions and the suitability of both \xrs\ and \proba/\lyra\ for statistical flare studies.  Finally, possible reasons for deviations from power law  distributions are discussed.  In Section~\ref{sec:obs} we discuss the \xrs\ and \proba/\lyra\ instruments and their flare detection algorithms.  In Section~\ref{sec:method} we describe our methodology, including the statistical tools used in analysing the data.  In Section~\ref{sec:results} we present the results and discuss their causes and implications.  Finally, in Section~\ref{sec:conc} we present our conclusions and outline what is required to determine the nature of the solar flare size distribution.

\section{Flare observations}
\label{sec:obs}
\subsection{Instrumentation}
\label{sec:instr}
\xrs\ \citep{hans1996} and \proba/\lyra\ \citep{sant2013, domi2013} both take spatially integrated full-disk measurements of the Sun.  The \xrs\ observes in two broadband channels -- 1-8~\AA\ (long) and 0.5-4~\AA\ (short) -- with a cadence of 3~s (2~s for \goes-15).  It is operated by the National Oceanographic and Atmospheric Administration (NOAA) for space weather monitoring.  Its near 100\% duty cycle and good calibration between satellites have caused it to become a standard candle to which more sophisticated scientific observations are typically compared.  Its longevity -- the first \xrs\ was launched in the  1970s -- has resulted in a nearly uninterrupted 40-year data set of flare observations and its use in numerous statistical flare studies \citep[e.g.\ ][]{lee1995, feld1997, shimo1999, vero2002a, batt2005, asch2012, ryan2012}.

\proba\ is a technology demonstration mission, which was launched in late 2009 and is operated from the Royal Observatory of Belgium \citep{zend2013}.  The \lyra\ instrument is comprised of three redundant units, each of which takes measurements in four broadband channels -- 120-123~nm (Lyman alpha), 190-220~nm (Herzberg), 17-80~nm~+~$<$~5~nm (Aluminium), and 6-20~nm~+~$<$~2~nm (zirconium) -- with a nominal cadence of 20~Hz.  However, cadence can be as high as 100~Hz during special campaigns.  \lyra's nominal unit (unit 2) typically has a near 100\% duty cycle making it another good candidate for monitoring flare activity.  However, its duty cycle drops between late September and early March when \proba\ experiences an eclipse season.  Since first light, \lyra's nominal unit has suffered severe degradation, particularly in the Lyman alpha and Herzberg channels, which lost almost all sensitivity within a year.  However, the aluminium and zirconium channels are far less degraded and their sensitivity to shorter wavelengths make them well suited to observing flares.

\begin{figure}
\begin{center}
\includegraphics[width=0.45\textwidth]{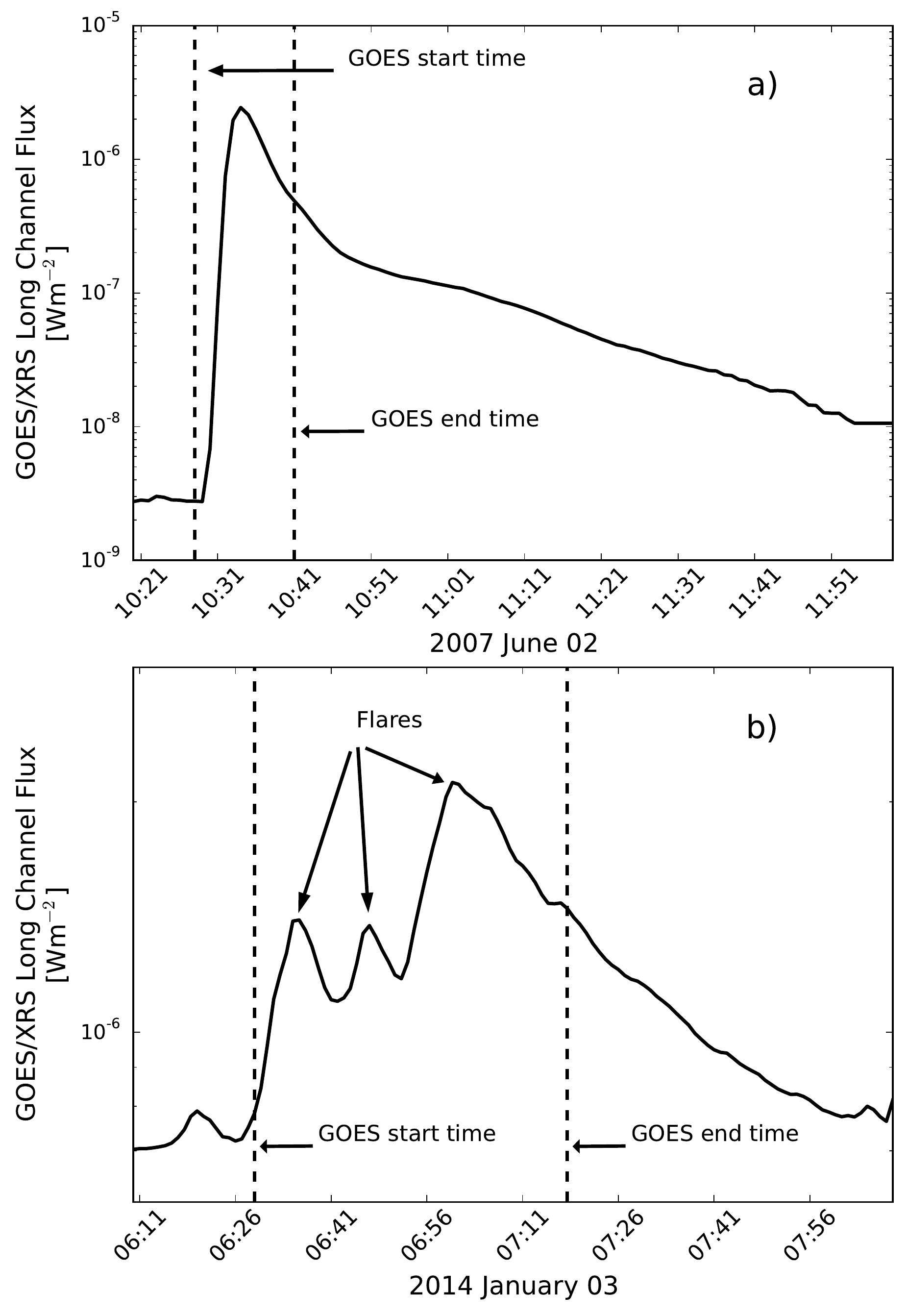}
\caption{a) \xrs\ one-minute-averaged lightcurve of a C2.4-class flare on 2007 June 03.  The vertical dashed lines mark the flare's start and end times according to the \goes\ event list.  It can be seen that the flare continues for over an hour after the the official \goes\ end time.  b) Example of three flares incorrectly labelled as one event by the \goes\ event list owing to crude flare definitions.}
\label{fig:gev_drawbacks}
\end{center}
\end{figure}

\subsection{ \goes\ event list}
\label{sec:gev}
Because of their suitability for monitoring and recording solar flare activity, automatic flare detection algorithms have been developed for both the \xrs\ and \proba/\lyra.  They serve as useful tools for space weather monitoring and valuable resources for scientific investigation.

The \goes\ event list flare detection algorithm is performed on one-minute-averaged  \xrs\ long channel data.  A flare is defined as having started if 
\begin{enumerate}
\item
there are four consecutive minutes of increasing flux;
\item
the flux at the end of the fourth minute is at least 40\% greater than the flux in the first minute.
\end{enumerate}
The flare peak is defined as the maximum flux observed during the flare, while a flare is defined as having ended once the flux drops to a value halfway between the initial and peak flux values, i.e. the time at which the flux equals $F_{peak} - (F_{peak} - F_{start}) \times 0.5$, where $F$ signifies flux.

These simple criteria make the \xrs\ flare detection algorithm easy and fast to implement and also fairly reliable for detecting larger flares.  This makes it well suited to space weather monitoring.  However, the criteria's simplicity can also cause difficulties for scientific analysis.  For example, the end time definition causes the flare duration to be systematically underestimated.  Therefore analyses of flare properties depending on the duration, e.g.\ fluence, is unreliable.  An example of this can be seen in Figure~\ref{fig:gev_drawbacks}a, which shows a \xrs\ lightcurve of a flare in the \goes\ event list with start and end times shown as dashed vertical lines.  It can be seen that the flare carries on long after the official end time.  An advantage of such an aggressive end time definition is that it enables the algorithm to better distinguish flares that occur temporally close together.  Since a new flare cannot be detected until the end of the previous one is found, a more realistic end threshold would cause flares occurring close in time to be counted as a single event, thereby corrupting the flare distribution.  In fact, even with this abrupt end time criterion, some events in the \goes\ event list are still made up of multiple flares.  An example can be seen in Figure~\ref{fig:gev_drawbacks}b, which shows three flares counted as one event.  In this case the flux of the first flare does not have time to reach the \goes\ event list end criterion before the second and then third flares dominate the signal.

It is difficult to determine exactly how often events like the one in Figure~\ref{fig:gev_drawbacks}b occur.  \citet{ryan2012} searched the \goes\ event list for instances during the decay phase of flares which met the \goes\ event list start criteria.  These secondary flares were not included in the \goes\ event list as the first flare had not yet met its end criterion.  They found that just over 5\% of listed events between 1991 and 2007 were `double flares', i.e.\ two (or more) flares listed as one.  This is a lower limit on the true number of double flares.  It does not take into account cases where the first flare is smaller than the second.  It only includes comparably sized double flares because smaller flares would not satisfy the \goes\ event list start criteria when placed on the elevated background level of the already decaying earlier flare.  It also does not account for flares with overlapping rise phases because they cannot be separated at all by the \goes\ event list.  Therefore, it is expected that double flares make up a small but significant fraction of events in the \goes\ event list.

\subsection{LYRA Flare Finder}
\label{sec:lyraff}
The LYRAFF algorithm (\lyra\ Flare Finder) was developed to enable \lyra\ to automatically detect flares for space weather monitoring as part of ESA's Space Situational Awareness programme (SSA).  The resulting flare list is available online\footnote{\url{http://proba2.oma.be/data/LYRA/LYRAFF}} and is updated in near-real time.  LYRAFF is very similar to the \goes\ event list, but has been designed to reduce the drawbacks described in Section~\ref{sec:gev}.  LYRAFF is applied to one-minute-averaged \lyra\ zirconium channel data.  This channel was chosen because it is sensitive to the shortest wavelengths and therefore exhibits the largest variability due to flares.  In addition this channel is the least degraded and so is more consistent over time.

\begin{figure}
\begin{center}
\includegraphics[width=0.45\textwidth]{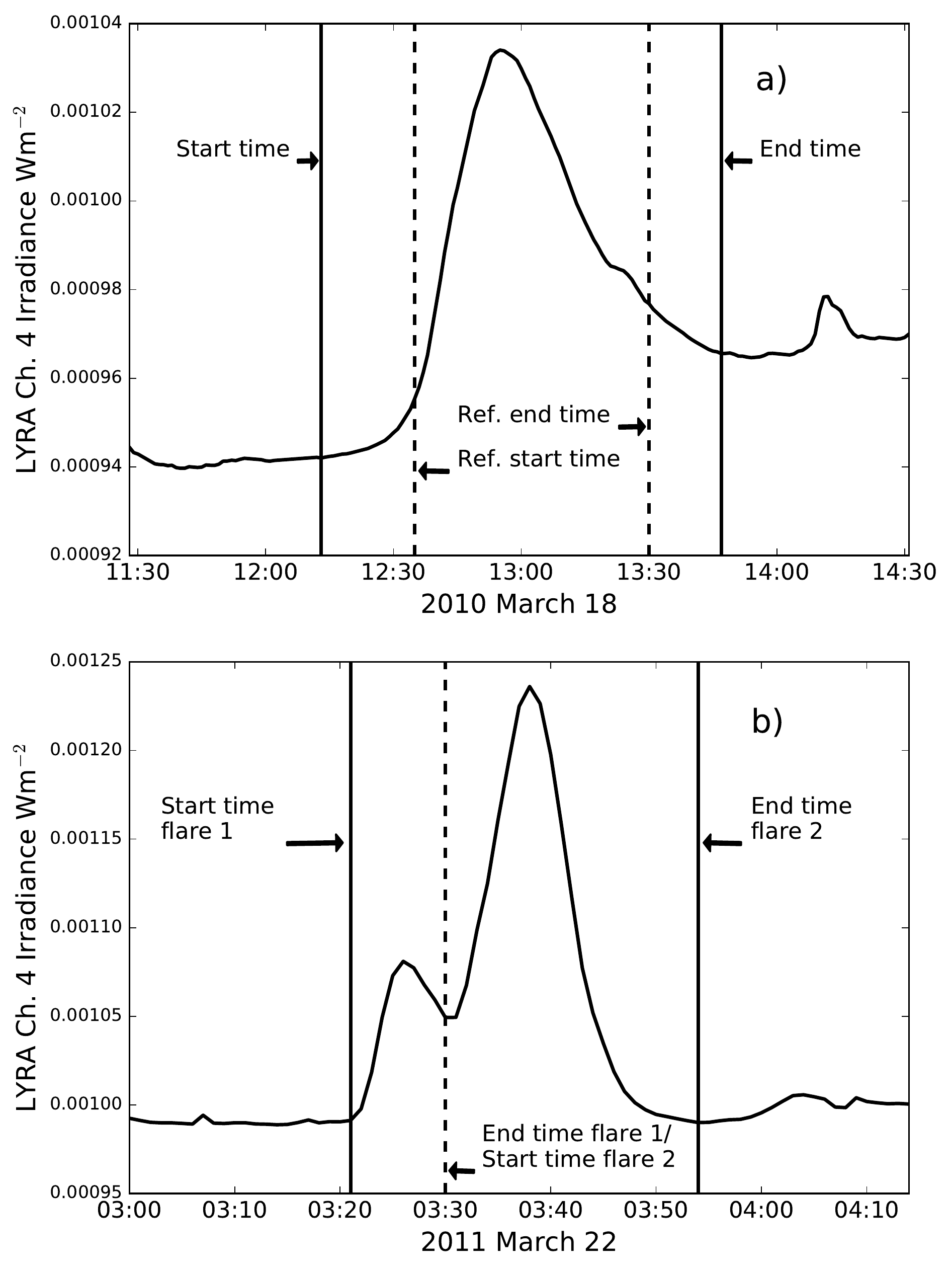}
\caption{a) \lyra\ zirconium one-minute-averaged lightcurve of a flare which occurred 2010 March 18.  The start and end times identified by the LYRAFF algorithm are marked by the solid vertical lines while the reference start and end times are marked by the dashed lines.
b) Example of two flares from 2011 March 22 which occurred temporally close together being successfully separated by LYRAFF.}
\label{fig:lyraff_example}
\end{center}
\end{figure}

Before flares can be sought, the numerous artefacts  often found in \lyra\ data must be removed (see Table~\ref{tab:lyra_artifacts} in Appendix~\ref{app:lyra_artifacts} for a list and explanation of these artefacts).  In addition, negative and zero irradiance values and those greater than or equal to 10\,Wm$^{-2}$ are deemed unphysical and owing to instrument error.  These data points are therefore also removed.  Once the data have been cleaned, LYRAFF identifies that a flare has started when
\begin{enumerate}
\item
four consecutive minutes of increasing irradiance are observed (without data gaps).  This is labelled the start threshold period;
\item
the irradiance in the fourth minute is 1\% greater than the irradiance in the first minute.  The first of these four minutes is labelled the reference start time, while the  1\% value is labelled the start threshold.
\end{enumerate}
Once the reference start time is found, LYRAFF better estimates the flare's true start time by tracking back to the first minute of continuously increasing flux, whether or not criterion 2) is met.  This is defined as the flare start time.  The flare peak is simply defined as the time of maximum irradiance.  Analogous to the reference start time, the reference end time is defined as the moment the irradiance drops to a level 50\% between the flare peak and flare start time irradiance values, just as it is for the \goes\ event list. The value of 50\% is labelled the end threshold. The flare end time is then found by tracking forward to the the time of the latest consecutive irradiance decrease.  By finding the earliest/latest times of consecutive increasing/decreasing irradiance, the estimates of the flare start/end times, and hence flare duration, are significantly improved.  This can be seen in Figure~\ref{fig:lyraff_example}a, which shows a \lyra\ zirconium channel lightcurve of a flare.  The reference start and end times are shown by the dashed vertical lines, while the official start and end times are shown by the solid vertical lines.  It is clear that the start and end times better approximate the flare duration than the reference times.

In order to reduce cases of double flares (multiple events being incorrectly labelled as a single event), LYRAFF simultaneously searches for a new reference start time once the irradiance has dropped 20\% from the peak towards the flare start irradiance.  This is labelled the first fall threshold.  If a new reference start time is found before the reference end time, the flare is defined as ending here and is added to the flare list and is  flagged as incomplete.  This technique improves the identification of flares which occur temporally close together.  Figure~\ref{fig:lyraff_example}b shows an example of two temporally close flares being successfully separated by the LYRAFF algorithm.  However, it is clear that in this case LYRAFF does not do a good job at approximating the flare duration.  Therefore, when examining properties related to duration, e.g.\ fluence, such events must be treated with caution.  Moreover, LYRAFF does not eliminate all cases of multiple flares being counted as one event.  If two flares happen simultaneously or temporally close enough together that the irradiance does not drop to the 20\% threshold, LYRAFF still cannot distinguish multiple events.

As a test of how effective LYRAFF is at detecting double flares, it was applied to the one-minute-averaged \xrs\ time series for 1991 to 2007 (inclusive) by altering the start threshold to 40\% (the \goes\ event list start threshold).  This period was chosen because it was the same that \citet{ryan2012} used to search for double flares in the \goes\ event list (Section~\ref{sec:gev}).  It was found that LYRAFF correctly identified and separated 799 double flares.  This was 2.6\% of all the flares listed in the \goes\ event list for this period and so is about half the percentage of double flares identified by \citet{ryan2012}.  Thus, LYRAFF significantly reduces the misidentification of double flares.

\section{Methodology}
\label{sec:method}
In order to explore how the choice of arbitrary thresholds in flare detection algorithms can affect the distributions they produce, we examined the variation in the flare peak flux frequency distribution.  As we discussed in Section~\ref{sec:intro}, such distributions are typically described as a power law.  We therefore determined the relationships between the power law slope of the background-subtracted peak flux size distribution and the algorithm threshold used to derive it.  If the power law slope of the flare peak distributions is independent of the choice of these thresholds, then statistical flare analyses based on these algorithms can be trusted.  However, if this is not the case, then the effect of threshold choice must be taken into account before drawing any conclusions about the statistical nature of flares.  In this section we outline the methodology followed in this study in three broad steps: producing flare lists using different algorithms and thresholds; performing a suitable background subtraction to the flare peak flux data; and fitting the flare peak flux frequency distributions with a power law.

\subsection{Producing Flare Lists}
\label{sec:flare_lists}
First we focussed on the \goes\ event list and \xrs\ data.  We used the one-minute-averaged long channel \xrs\ data produced by NOAA\footnote{\url{http://satdat.ngdc.noaa.gov/sem/goes/data/new_avg/}} from 1986 to 2014, inclusive.  Data points marked as bad in the data files were removed, as were negative and unphysically high flux values.  We then ran the \goes\ event list algorithm several times for values of the start threshold in the range 5\%--90\%\footnote{For clarity, a 5\% end threshold means the flare ends when the flux has drops to a level equal to $F_{peak} - (F_{peak}-F_{start}) \times 0.05$ where $F$ is flux.  Thus, a 5\% end threshold results in a short flare duration, while a 90\% end threshold results in a long flare duration.} while keeping the end threshold fixed at its standard value of 50\%.  We then repeated the process for different values of end threshold in the range 5\%--90\% while keeping the start threshold fixed at its standard value of 40\%.  Thus two sets of \goes\ flare lists were produced reflecting the effect of the start and end thresholds.

Next we turned to LYRAFF and \proba/\lyra.  We constructed a one-minute-averaged \lyra\ zirconium channel time series from January 2010 to May 2015 inclusive.  The time series was cleaned as outlined in Section~\ref{sec:lyraff}.  Eclipse season observations (mid-September to mid-March) were also removed as attenuation by the atmosphere causes the irradiance signal to smoothly but dramatically rise and fall over the course of each orbit, making it impossible for LYRAFF to reliably detect flares.  LYRAFF was then run several times on this time series using different values of start threshold in the range 0.1\%--100\% while keeping the first fall and end thresholds fixed at their standard values (20\% and 50\%, respectively).  The process was repeated for different values of the first fall threshold in the range 5\%--90\%, while keeping the start and end thresholds fixed at their standard values (1\% and 50\%).  The process was repeated once more for different values of end threshold in the range 5\%--90\%, while keeping the start and first fall thresholds fixed at their standard values (1\% and 20\%).  This resulted in three sets of \lyra\ flare lists corresponding to varying start, first fall, and end thresholds, respectively.

Finally, we also applied LYRAFF to the  \xrs\ time series.  We produced three sets of flare lists where one LYRAFF threshold was varied in the range 5\%--90\%, while the other two were held fixed at their standard values (40\%, 20\%, and 50\% for the start, first fall, and end thresholds, respectively).  Thus we produced eight sets of flare lists in total, each allowing us to explore the effect of a certain threshold.

\subsection{Background subtraction}
\label{sec:bgsub}
Before the flare peak flux size distributions could be reliably analysed, the background flux had to be subtracted.  Otherwise, the derived distributions would be unphysically steep.  Since the rise phase of many flares is of the order of minutes, the background typically has little chance to evolve between the start and peak of the flare.  Therefore in many cases the flux at the flare start adequately approximates the background at the peak time.  For some flares though, such as those occurring on the tails of other flares, this will not be the case.   However,  as the \goes\ event list does not reliably determine the flare end time, more sophisticated methods of determining the background -- such as fitting a polynomial to periods of pre-flare and post-flare flux -- are also likely to be unrepresentative of the true background.  For flare distributions derived using LYRAFF, which does a better job of determining the flare duration, we compared the results of subtracting the initial flux and a value at the peak time found from linear interpolation between the start and end of the flare.  However, we found no significant difference in the subsequent results.  We therefore simply subtracted the initial flare flux for all cases.

\subsection{Fitting flare distributions}
\label{sec:fitting}
\begin{figure}
\begin{center}
\includegraphics[width=0.45\textwidth]{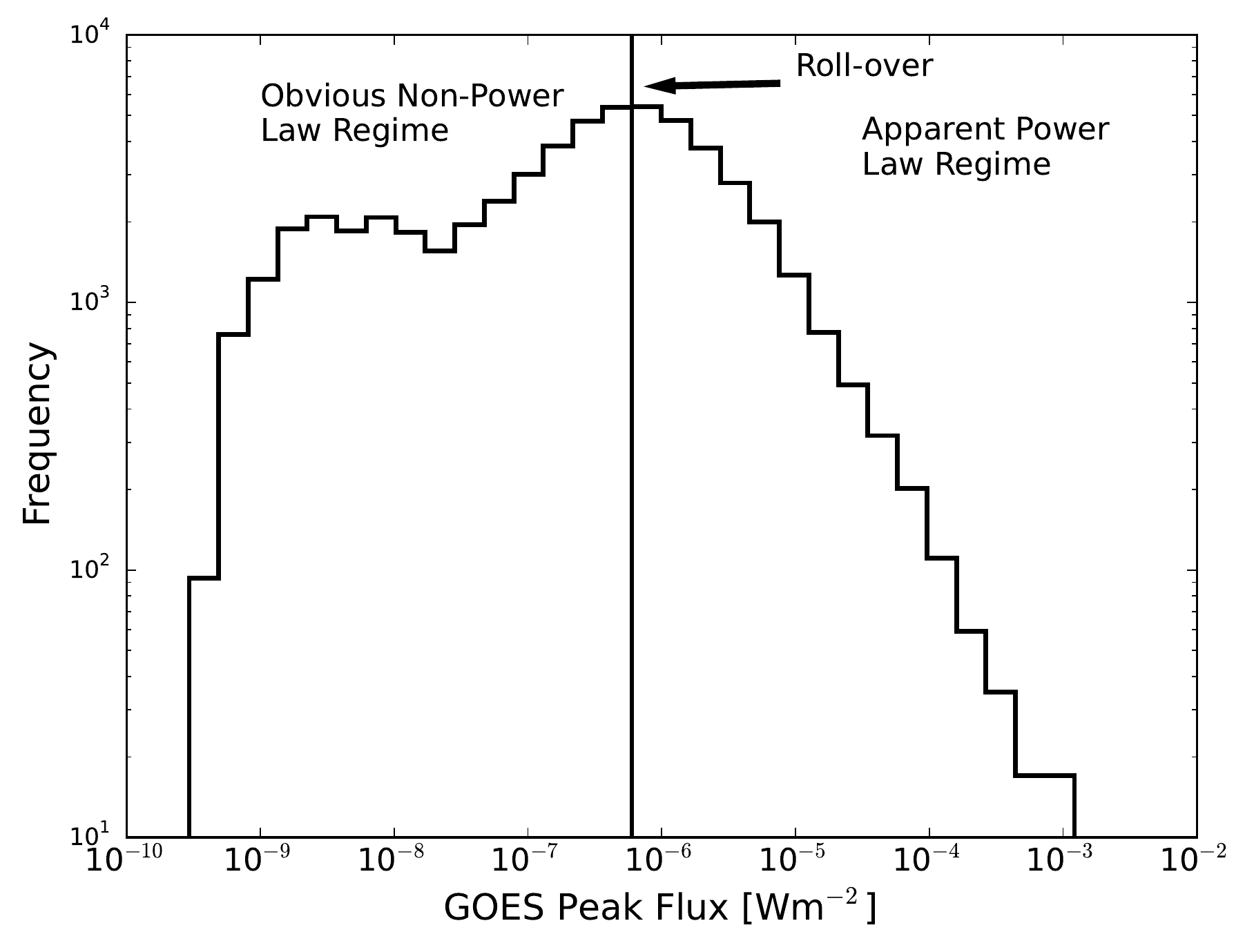}
\caption{Frequency distribution of the background-subtracted \goes\ long channel peak flux for flares detected using the standard \goes\ event list algorithm (start threshold=40\%, end threshold=50\%).  Above the roll-over the distribution appears to follow a power law, while below the roll-over it deviates owing to undersampling of the true flare distribution.}
\label{fig:example_gev_hist}
\end{center}
\end{figure}

Figure~\ref{fig:example_gev_hist} shows a histogram of the background-subtracted flare peak flux size distribution corresponding to the standard \goes\ event list (start threshold=40\%, end threshold=50\%).  It can be seen that above $\sim$10$^{-6}$\,Wm$^{-2}$ the distribution appears be a power law.  The roll-over below this level is caused by undersampling of events owing to the variation in the solar background, which typically reaches low C-class at solar maximum.  All \goes\ flare peak flux frequency distributions derived in this study exhibited this same qualitative behaviour.  

In order to quantitatively determine the effect of the different flare algorithm thresholds, we fit each distribution with a power law and examined the variation of the exponent.  To do this we used maximum likelihood estimation (MLE), which is a standard statistical technique for fitting parametric models to empirical data.  It is far more robust and easy to use than graphical methods such as fitting a linear model to a log-log histogram.  For further discussion of the advantages of MLE see \citet{clau2009} and specifically in the context of solar physics, see \citet{dhuy2016}.

Given a parametric model (e.g.\ a power law) described by a parameter, $\alpha$, and a set of observations of a property, $x$, it is possible to determine the likelihood function, $L(\alpha, x)$, which describes the probability that the data is drawn from a distribution with a given $\alpha$ value.  According to MLE, the maximum of this likelihood function gives the most likely estimate of $\alpha$ (denoted $\hat{\alpha}$), i.e.\ the best fit.  This can simply be determined by differentiating the likelihood function, setting it to zero, and solving for $\alpha$.  Therefore, given a power law of the form
\begin{equation}
\label{eqn:powerlaw}
f(x) \propto x^{-\alpha}
\end{equation}
the most likely estimate of $\alpha$ is given by
\begin{equation}
\label{eqn:mle_powerlaw}
\hat{\alpha} = 1 + n_{tail} \left( \sum^{n_{tail}}_{i=1} ln \frac{x_i}{x_{min}}\right)^{-1}
,\end{equation}
where \xmin\ is the lower limit above which the data obey a power law (e.g.\ the roll-over in Figure~\ref{fig:example_gev_hist}); $x_{i=1,...,n_{tail}}$ is the subsample of data, where $x_i \geq x_{min}$; and $n_{tail}$ is the number of data points in the subsample $x_{i=1,...,n_{tail}}$.  To a first-order approximation, the statistical error on $\hat{\alpha}$ is then given by
\begin{equation}
\label{eqn:mle_powerlaw_uncert}
\sigma = \frac{\hat{\alpha}-1}{\sqrt{n_{tail}}}
.\end{equation}
For a derivation of Equations~\ref{eqn:mle_powerlaw} and \ref{eqn:mle_powerlaw_uncert}, see Appendix B of \citet{clau2009}.

In order to obtain a reliable fit via MLE, a good estimation of \xmin\ must first be made.  There are a number of ways of doing this.  It can be approximated by eye from histograms.  However, this depends on the arbitrary choice of bin width and the scientist's subjective choice of where the histogram deviates from a power law.  Therefore, in order to determine the lower limit more robustly, we performed the MLE power law fit for \xmin\ equal to each flux value in the sample.  We then selected the \xmin\ which produced the lowest Kolmogorov-Smirnov (KS) statistic\footnote{The KS statistic is a standard measure of how well an empirical distribution is described by a fit.  It is given by the maximum difference between the empirical cumulative distribution function (CDF) of the data and the theoretical CDF of the fit.  The smaller the KS-statistic, the more closely the fit approximates the data.}.  Using this method we fit power laws to the peak flux/irradiance frequency distributions of all the flare lists derived in Sections~\ref{sec:flare_lists} and \ref{sec:bgsub}.  

It should be pointed out that MLE is not the only viable fitting technique available.  Other studies \citep[e.g.][]{irel2015} have used Markov chain Monte Carlo (MCMC) methods to fit power laws to solar observational data.  The advantages of MCMC methods are that they provide a more rigorous determination of the fit uncertainties and tend to be more robust in analysing complex likelihood functions that may include numerous local maxima.  However, for well-behaved likelihood functions, MLE is expected to produce similar results to MCMC.  What is more, MCMC methods come at an additional computational cost.  In their analysis of the Fourier power spectra of the dynamics of different regions of the solar atmosphere, \citet{irel2015} ran each MCMC chain for 150,000 samples. In our case, implementing this procedure for over 100 separate data sets would be prohibitive.  We therefore chose MLE as our preferred fitting technique for this study.  In the next section we discuss the results of our analysis.

\section{Results and discussion}
\label{sec:results}

\begin{figure*}
\begin{center}
\includegraphics[width=0.7\textwidth]{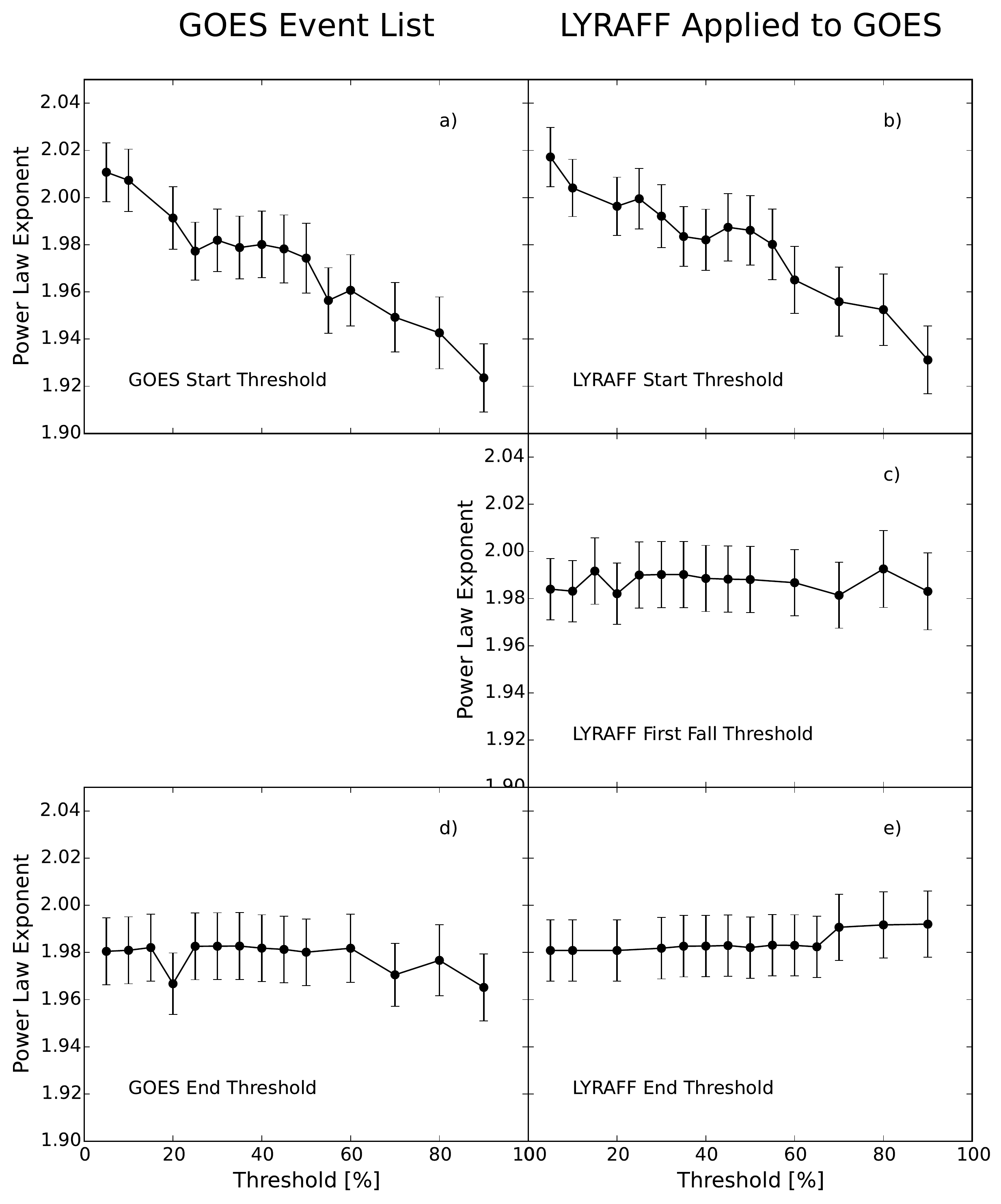}
\caption{Power law slope of the background-subtracted \goes\ long channel peak flux frequency distribution as a function of the \goes\ event list and LYRAFF thresholds.  The left column corresponds to the thresholds of the \goes\ event list algorithm, while the right column corresponds to the LYRAFF algorithm applied to \xrs\ observations.  Each row then corresponds to the start (top), first fall (middle), and end (bottom) thresholds.  We note that there is no plot in the middle left panel as the \goes\ event list does not use a first fall threshold.  The error bars represent the statistical uncertainty of the power law fits given by Equation~\ref{eqn:mle_powerlaw_uncert}.  We note that there is a clear downward trend in both start threshold plots implying that the derived flare distribution depends on the choice of start threshold.  In contrast, the first fall and end threshold panels show flat relationships implying that the derived flare distribution is independent of the choice of these thresholds.}
\label{fig:goes_exp_vs_thresh}
\end{center}
\end{figure*}

\subsection{ Effect of arbitrary thresholds on \goes-derived flare distributions}
\label{sec:goes_thresholds}
Figure~\ref{fig:goes_exp_vs_thresh} shows the relationships between the thresholds used to define \goes\ flares and the power law exponent of the derived \goes\ flare peak flux size distribution.  The power law exponents of each distribution were obtained using \xmin\ in the range 5$\pm$1\,$\times$\,10$^{-6}$\,Wm$^{-2}$.  The mathematical justification for this choice is that an \xmin\ in this range gives the lowest KS statistic.  The physical justification is that this is the maximum level  the solar X-ray background can reach during solar max.  This causes the flare detection algorithms to undersample the true flare distribution, thereby making the statistics unreliable below this level.

Figure~\ref{fig:goes_exp_vs_thresh}a shows the variation of the power law exponent with the \goes\ event list start threshold.  A shallow but clear decreasing dependence is evident.  For a start threshold of 5\% the exponent is greater than 2.  As the threshold is increased the exponent decreases steadily until it reaches $\sim$1.92 for start thresholds of 90\%.  Moreover this trend is clearly beyond the range of statistical uncertainty.  We might naively expect this result if we used crude fitting techniques and believed that the change in start threshold simply caused the detection algorithm to undersample the flare population below a given peak flux.  This would only cause the lower section of the distribution to deviate from a power law and would hence flatten the average slope.  However, by robustly determining $x_{min}$ and excluding events below this value, we would expect the fitted power law slope not to change as the larger flares should be common to all flare distributions.  However, this is not what we see in Figure~\ref{fig:goes_exp_vs_thresh}a.  We see that despite robustly determining $x_{min}$, the start threshold still affects the power law slope.  This occurs because the start threshold defines a flare based on its impulsiveness rather than its peak flux.  Therefore, some larger flares above the best found value of $x_{min}$ may occasionally be missed because they are not impulsive enough.  This shows that -- depending on how the start of a flare is defined -- different flare distributions can be obtained.  

To verify this result, we examined the same relationship for the LYRAFF algorithm (Figure~\ref{fig:goes_exp_vs_thresh}b).  The improved ability of LYRAFF to detect temporally nearby flares helps determine whether the above result is owing to incorrect classifications of multiple events or aggressive start and end time definitions.  However, a very similar behaviour is observed.  Although the decreasing trends in Figures~\ref{fig:goes_exp_vs_thresh}a and \ref{fig:goes_exp_vs_thresh}b are small, they could be significant.  For example, when trying to determine whether nanoflaring can explain the coronal heating problem by extrapolating the flare distribution, the key value of the power law slope is 2.  If we naively define a flare with a 5\% or 50\% start threshold, we could come to totally opposite conclusions.  Therefore, it is vital to fully understand the effect of the chosen flare definitions before interpreting the results.

Despite the similarity between Figures~\ref{fig:goes_exp_vs_thresh}a and \ref{fig:goes_exp_vs_thresh}b, the results are not exactly the same.  With the exception of the 10\% start threshold, the LYRAFF exponents are all greater than their \goes\ event list counterparts for the same threshold by between 0.1\% and 1.2\%.  This is a very small difference and within the range of uncertainty.  Nonetheless, a slightly higher exponent is expected when employing LYRAFF.  As discussed in Section~\ref{sec:lyraff},  LYRAFF successfully identified and separated 2.5\% of the flares in the standard \goes\ event list as double flares, using the standard \goes\ thresholds (start threshold=40\%, first fall threshold=20\%, end threshold=50\%).  These additional flares are the main difference between the flare lists derived by applying the \goes\ event list and LYRAFF algorithms to \xrs\ time series.  The peak flux frequency distribution of these additional flares has a best fit power law exponent of 2.59$\pm$0.15 above the $x_{min}$ value of 5$\times$10$^{-6}$\, Wm$^{-2}$ commonly used in this study.  This is significantly steeper than the corresponding LYRAFF exponent in Figure~\ref{fig:goes_exp_vs_thresh}b as the distribution of the additional flares is highly skewed to smaller events.  This occurs because it is the smaller flare that is masked from the peak flux size distribution when a double flare occurs.  Therefore, LYRAFF's ability to better separate double flares is what leads to marginally steeper flare distributions.  However, as also highlighted in Section~\ref{sec:lyraff}, LYRAFF cannot identify all temporally overlapping flares.  If more of these flares could be identified, it would only serve to further steepen the overall flare size distribution. This shows the small but important impact that correctly identifying temporally overlapping flares can have.

Next, we examined the effect of the first fall and end thresholds of the \goes\ event list and LYRAFF algorithms, seen in Figure panels \ref{fig:goes_exp_vs_thresh}c--\ref{fig:goes_exp_vs_thresh}e.  More promisingly, flat relationships around 1.98--1.99 were found.  This demonstrates that the derived flare distribution is not influenced by the choice of first fall or end thresholds in either algorithm.  However, it must be noted that we are only looking at peak flux distributions which are independent of flare duration.  If we were to examine fluence, the choice of first fall and end thresholds and the effectiveness of our algorithm to accurately determine start and end times would play a very important role.

\begin{figure}
\begin{center}
\includegraphics[width=0.4\textwidth]{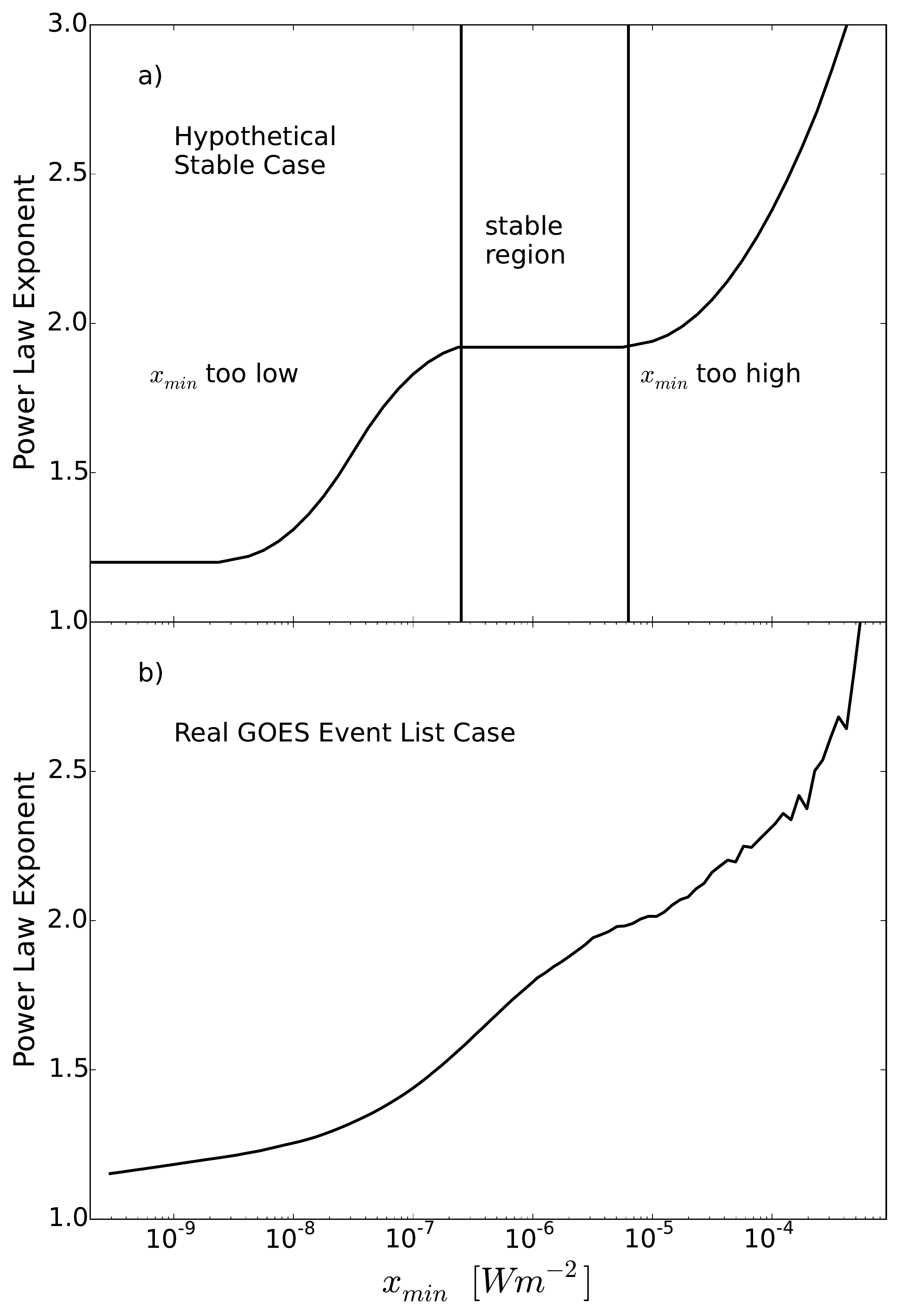}
\caption{a) Hypothetical relationship between the best fit power law slope and the lower limit (\xmin) for a stable power law fit.  The flat region in the middle of the plot shows the range over which the fit is independent of the choice of \xmin, i.e.\ stable.  b) The same relationship as a) for the real \goes\ long channel peak flux frequency distribution derived with the standard \goes\ event list (start threshold=40\%, end threshold=50\%).  It can be seen that there is no flat region where the power law slope is independent of choice of \xmin.  This suggests that the distribution may not be best approximated by a single power law.}
\label{fig:gev_stability}
\end{center}
\end{figure}

\subsection{ Stability of the power law nature of flare distributions}
\label{sec:stability}
\subsubsection{Stability plots}
\label{sec:stability_plots}
In order to further investigate the reliability of the power law fits discussed in Section~\ref{sec:goes_thresholds}, we examined the stability of the MLE-derived power law exponent as a function of \xmin.  Figure~\ref{fig:gev_stability}a shows a cartoon of a stability plot,  i.e.\ the dependence of the derived fit on the choice of \xmin,  for a hypothetical stable power law fit \citep[for similar plots, see][]{clau2009}.  There are three regimes in a stability plot for a reliable MLE-derived fit.  At low values of \xmin\ there is an increasing relationship  because \xmin\ is lower than the roll-over (Figure~\ref{fig:example_gev_hist}) caused by the undersampling of events.  As \xmin\ is decreased below the roll-over, the fit is made artificially shallower.  At high \xmin\ there is another steeper increasing relationship because the choice of \xmin\ is now so high that we only have a few events in our sample which is insufficient for fitting a power law.  At intermediate values of \xmin\ the relationship is flat.  This means that within this range the derived power law exponent is independent of the choice of \xmin\ since the distribution is now adequately sampled.  However, as the choice of \xmin\ is increased, the number of events in the sample is decreased, causing the statistical uncertainty of the fit to increase.  Therefore, it is important to choose the lowest value of \xmin\ for which the fit is stable.  If the dependence of the derived power law exponent on \xmin\ exhibits this flat regime, we say that the power law fit is stable.

Figure~\ref{fig:gev_stability}b shows the stability plot for the \goes\ peak long channel flux size distribution derived using the standard \goes\ event list definitions (start threshold=40\%, end threshold=50\%).  It can be seen that the intermediate flat regime is absent.  The stability plots of all the derived flare peak flux size distributions in this study show the same qualitative behaviour.  This implies that these distributions are not well fit by a single power law.  This is in contrast to numerous studies over the past   decades which have treated the distribution of peak \goes\ long channel flux as such.

\subsubsection{Assessing goodness-of-fit}
\label{sec:p-values}
To better understand this result, we assessed the goodness-of-fit for all the flare data sets derived in this study to their respective power law fits. The methods described below were implemented with the \textbf{poweRlaw} package \citep{gill2015} using the statistical software language R.  We initially operated under the assumption that the observed data greater than $x_{\min}$ follows a power law distribution  (our null hypothesis).  We then use $p$-values to assess the plausibility of this assumption in each case. The $p$-value is the probability that a finite sample drawn from a parent distribution will deviate from that parent distribution by at least a given amount due to random chance. In our case, the finite sample is the observed peak flux frequency distribution, while the believed parent distribution is a power law. If $p$ is large, then it is impossible to rule out random noise as the cause of the deviation of the data from a power law. If $p$ is small, then there is a very low probability of this being the case and our assumption that the null hypothesis is true becomes untenable.  (For a description of how $p$-values were calculated in this study, see Appendix~\ref{app:p}.)

Figure~\ref{fig:goes_pvalues} shows a histogram of the $p$-values derived for all the \xrs\ flare lists derived with different thresholds in the \goes\ event list (solid) and LYRAFF (dashed) algorithms.  The vertical dotted line shows  $p = 0.05,$ which is a threshold typically used to define whether a $p$-value is significant.  It can clearly be seen that the majority of flare lists have $p$-values less than 0.05, while almost all have a $p$-value less than 0.1.  This means that almost all flare distributions derived in this study have less than a 1 in 20 chance that their deviation from a power law is due to random noise.  Moreover, almost all have less than a 1 in 10 chance.  Although this does not categorically prove that the distributions are not power laws, it does cast significant doubt over our above assumption that flares detected with \xrs\ are appropriately modelled by a power law parent distribution.

\begin{figure}
\begin{center}
\includegraphics[width=0.5\textwidth]{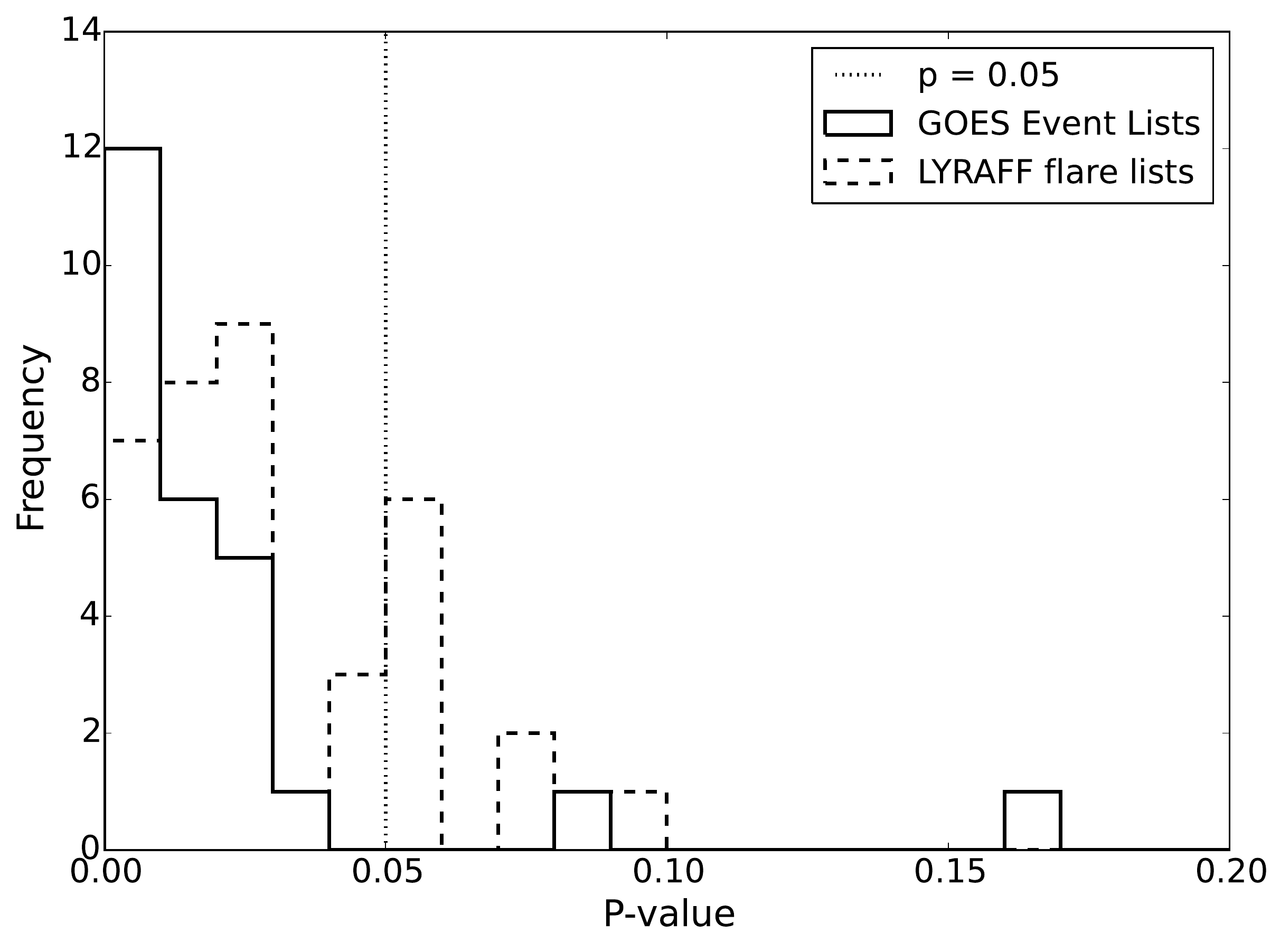}
\caption{Histogram of p-values for power law fits to all \xrs\ flare peak irradiance frequency distributions determined with the \goes\ event list (solid) and LYRAFF (dashed).  The dotted vertical line denotes the 5\% level below which we reject the null hypothesis, i.e.\ that the flares are drawn from a power law distribution.}
\label{fig:goes_pvalues}
\end{center}
\end{figure}

\subsubsection{Possible Causes Of Non-Power Law Behaviour}
\label{sec:stability_causes}
There are two different   explanations for why the observed flare peak flux size distributions are not  power law-like.  Either the true flare distribution is power law-like and the observational and/or analytical techniques are masking this  or the true flare distribution is in fact not a power law.

There are a few possible reasons why this type of analysis may not reveal a power law flare distribution.  First, the simple background subtraction methods used here --  subtracting the flux at the flare start time or a linearly extrapolated flux between the flare start and end times -- may not be adequate and may be corrupting the distribution.  It is difficult to fully rule this out without performing manual background subtraction for all events and comparing results.  Second, the range over which we can fit a power law may be insufficient to adequately determine the power law stability.  The \xmin\ values used in the fits described in Section~\ref{sec:fitting} are around 5$\times$10$^{-6}$\,Wm$^{-2}$ owing to a combination of the upper limit of solar background and the \xrs\  lack of spatial resolution.  Moreover, the largest flares in our sample are of the order of 10$^{-3}$\,Wm$^{-2}$, while we have only very small number statistics above 10$^{-4}$\,Wm$^{-2}$.  This gives us less than two orders of magnitude over which to fit a power law.  If we were able to extend this range, perhaps a more reliable, stable power law could be found.  Third, even in the range 10$^{-6}$--10$^{-4}$\,Wm$^{-2}$ we could still be undersampling the true flare distribution owing to large flares masking smaller ones.

In order to investigate whether this third possibility is plausible, we attempted to determine how many flares would have to be missed by \xrs\ and/or the standard \goes\ event list algorithm (start threshold=40\%, end threshold=50\%) to make the observed flare frequency distribution a stable power law.  To do this we added synthetic flares to the original distribution above the upper limit of solar background variation ($\sim$5$\times$10$^{-6}$\,Wm$^{-2}$) until the CDF of the flare distribution closely matched the CDF of the theoretical power law distribution.  The difficulty in this method is selecting the power law exponent with which to generate the theoretical CDF.  Since the stability plot in Figure~\ref{fig:gev_stability} is continuously increasing there is very little reason to select one exponent value over another.  On the one hand, exponents corresponding to an \xmin\ close to the upper limit of the solar background are, according to our hypothesis, artificially shallow because of undersampling.  On the other, exponents corresponding to a high \xmin\ suffer from small number statistics and are therefore artificially steep.  To give us an idea of the importance of this problem, we performed this process for exponents of 1.98, 2.02, 2.2, and 2.3  (which correspond to fits using \xmin\ values of 5$\times$10$^{-6}$, 10$^{-5}$, 5$\times$10$^{-5}$, and 10$^{-4}$\,Wm$^{-2}$, respectively; see Figure~\ref{fig:gev_stability}).  The number of additional flares required to make the distribution stable around an \xmin\ of $\sim$5$\times$10$^{-6}$\,Wm$^{-2}$ for each of the exponents were found to be as follows: an exponent of 1.98 required $\sim$300--400 additional flares; an exponent of 2.02 required $\sim$700--800 additional flares; an exponent of 2.2 required $\sim$2500--4000 additional flares; and an exponent of 2.3 required $\sim$4000--6000 additional flares.  As percentages of the observed number of flares with peak fluxes above 5$\times$10$^{-6}$\,Wm$^{-2}$, these numbers translate to 6--8\%, 14--16\%, 25--80\%, and 80--120\%.  The distributions of additional flares were predominantly high C- and low M-class flares.  The percentage corresponding to an exponent of 1.98 is very similar to the 5\% of flares in the \goes\ event list found by \citet{ryan2012} to be double flares. As discussed in Section~\ref{sec:gev}, this is a lower limit for the number `missing flares' from the \goes\ event list, so perhaps the true number may be as high as $\sim$15\%.  This would agree with the 14--16\% of additionally required flares corresponding to a power law exponent of 2.02.  Therefore if the true power law exponent of the flare size distribution is around 2, it is not impossible that the deviation from a power law is due to undersampling of high C- and low M-class flares.  If the true power law exponent is much above 2, this becomes increasingly unlikely.  Unfortunately, however, it is not possible to objectively determine the true exponent from the \xrs\ observations alone.

Other possible explanations for the behaviour of the observed flare frequency distribution is that the true distribution is close to but not actually a power law.  If the global flare distribution is a convolution of multiple heavy-tailed distributions,  e.g.\ if each active region has a different flaring distribution or if there are different flaring mechanisms which produce different distributions,  then the resulting distribution would converge to a power law under the central limit theorem.  Evidence for active region flare distributions deviating from a power law was found by \citet{whea2010}.  He used \xrs\ to examine AR 11029, which produced 70 B- and C-class flares.  Using a Bayesian maximum likelihood estimation technique he found that a power law plus exponential rollover model was 200  times more likely to represent the data than a simple power law.  This functional form qualitatively agrees with the unstable single power law fits found in this study.  \citet{whea2010} interpreted this power law deviation as being due to the fact that the active region only had a finite amount of free magnetic energy available for flaring.  If this behaviour is found for all or many active regions, it is possible that the convolution of all active region flare distributions studied here would also exhibit this behaviour.

Another possibility is related to the fact that traditionally the \goes\ flare peak flux has been used as a proxy for the flare energy.  This is valid if the total flare energy scales reliably with \goes\ peak flux.  However, this may not be the case.  It may be possible that some flares have a lower proportion of their thermal emission in the \xrs\ long passband owing to their temperature or differential emission measure (DEM) distribution.  This may cause the flare distribution as observed by \xrs\ to slowly flatten at lower peak fluxes.  Previous studies have examined the effect of using X-ray and EUV flux measurements as a proxy for the total flare energy \citep{asch2002,pare2006}.  These studies have found that the power law slope of the flare size distribution as observed with a given emission line/passband can be different from the size distribution of flare thermal energies which ultimately drives the observed emission.  This difference was found to depend on the formation temperature/temperature response of the observed line/passband.  \citet{pare2006} further concluded that high temperature lines/passbands tend to be less different from the thermal energy size distributions.  These studies, however, did not focus explicitly on the validity of a power law model for the size distribution, but rather the best fit power law slopes.

The interplay between conductive and radiative cooling is another potentially important factor.  This interplay varies depending on the flare temperature.  The \xrs\ temperature response lies in the range 4--40\,MK and  typically leads it to  record flare peak temperatures in the range 10--25\,MK.  At these temperatures, conduction is more efficient than radiative cooling, with radiation becoming increasingly relatively efficient as the temperature drops \citep{carg1995}.  The cooling rates change from flare to flare based on physical conditions \citep{ryan2013}.  Therefore, a greater or lesser fraction of a flare's thermal energy may be transported through conductive or radiative processes and hence emitted in the \xrs\ passbands at the flare peak.

Furthermore, flare energy may not be consistently partitioned between thermal energy, accelerated particles, non-thermal emission, etc.  
A comprehensive understanding of the energy partition between the different physical processes is a large and difficult undertaking and is beyond the scope of this paper.  There have been some previous studies which have made important progress in this endeavour \citep[e.g.\ ][]{emsl2012,asch2015}.  However, more work is required to fully understand how the energy partition changes as a function of peak \goes\ flux across multiple orders of magnitude.

\begin{figure}
\begin{center}
\includegraphics[width=0.5\textwidth]{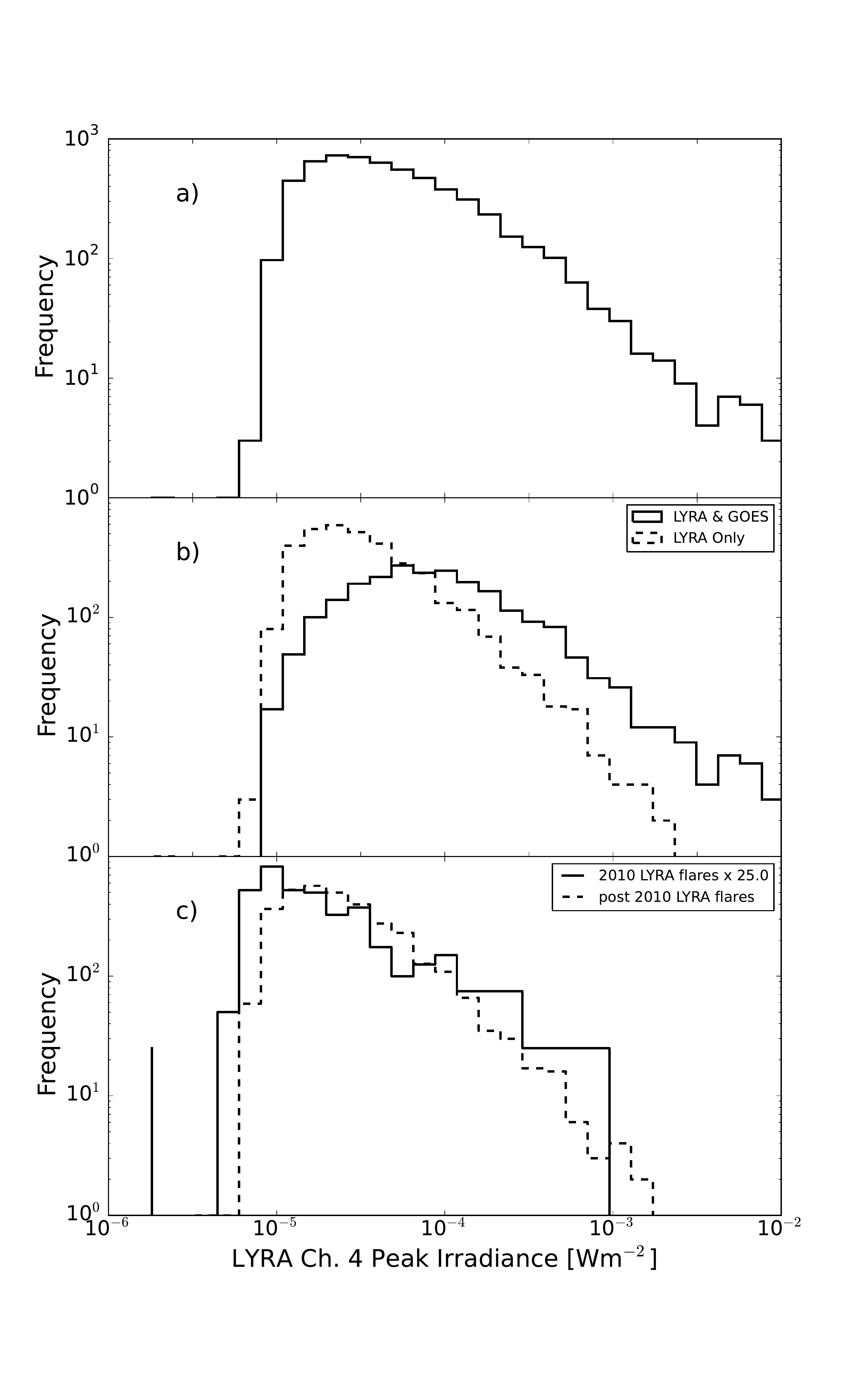}
\caption{a) Frequency distribution of background-subtracted \lyra\ zirconium channel peak irradiance of all non-eclipse-season flares between January 2010 and May 2015 derived with the standard LYRAFF algorithm (start threshold=1\%, first fall threshold=20\%, end threshold=50\%).  The distribution above the roll-over does not follow a straight line indicating this distribution is not a power law.  b) The same distribution as in a) broken into flares observed by \lyra\ and \xrs\ (solid) and those observed by just \lyra\ (dashed).  The flares observed by only \lyra\ tend to be smaller and have a steeper distribution. c) The same distribution as in a) broken into flares which occurred in 2010 and after 2010.  The slope of the 2010 distribution is less steep.  We note that  the 2010 distribution has been scaled by 25 to make it easier to visually compare its slope with the post 2010 distribution.}
\label{fig:lyra_hists}
\end{center}
\end{figure}

\subsection{Using \lyra\ for flare statistics}
\label{sec:lyra_flare_stats}
\begin{figure}
\begin{center}
\includegraphics[width=0.45\textwidth]{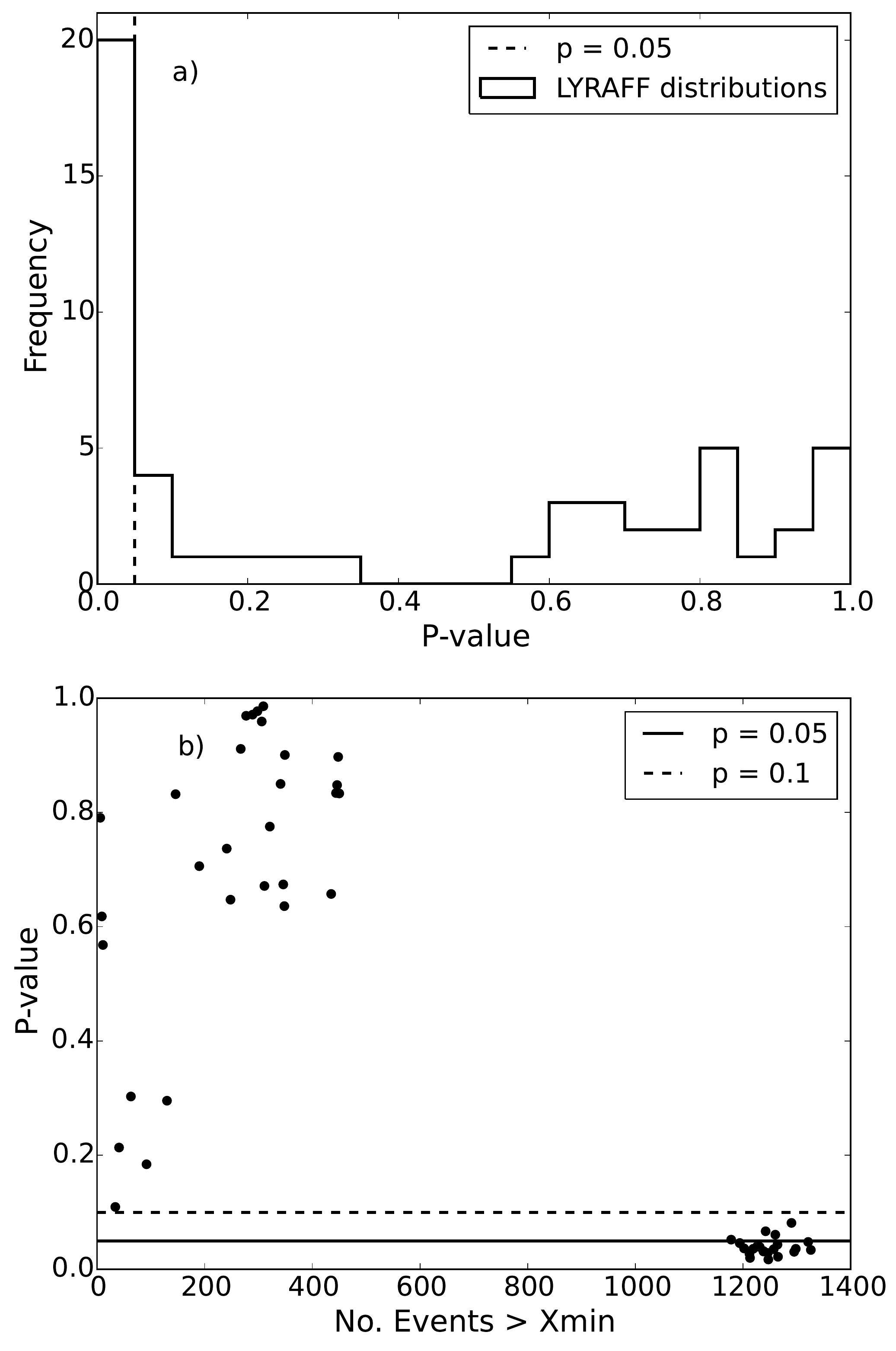}
\caption{a) Histogram of p-values for power law fits to all \lyra\ flare peak irradiance size distributions determined with the LYRAFF.  The dashed vertical line denotes the 5\% level below which we reject the null hypothesis, i.e.\ that the flares are drawn from a power law distribution.  b) p-value vs. number of flares greater than \xmin\ for the same flare distributions as in a).  We note that distributions with high p-values have less than 500 events above \xmin, causing the power law fits to be unreliable.}
\label{fig:lyra_pvalues}
\end{center}
\end{figure}

We performed a similar analysis as described in Section~\ref{sec:goes_thresholds} on the \lyra\ flare lists derived with LYRAFF.  Figure~\ref{fig:lyra_hists}a shows a log-log histogram of the background-subtracted peak \lyra\ zirconium channel irradiance of flares detected with the standard LYRAFF thresholds (start threshold=1\%, first fall threshold=20\%, end threshold=50\%).  The first thing to note is that the distribution visibly deviates from a straight line even above the roll-over.    Similar behaviour was observed in most of the \lyra\ peak irradiance size distributions derived with different LYRAFF thresholds.  This implies more strongly than in the case of \goes\ that these distributions are not well described by a power law.  To confirm this, we determined the p-values of each \lyra\ peak irradiance frequency distribution using the same process described in Section~\ref{sec:stability}.  The results are shown in Figure~\ref{fig:lyra_pvalues}a.  It can be seen that most of the distributions have a p-value less than 0.05, which suggests that our null hypothesis -- that the flares are drawn from a parent power law distribution -- is incorrect.  This agrees with our visual observation from Figure~\ref{fig:lyra_hists}a.  However, there is also a minority of distributions with very high p-values.  We investigate this in Figure~\ref{fig:lyra_pvalues}b, which shows the p-values as a function of the number of flares used in the fitting process, i.e.\ the number of flares greater than \xmin.  It can be seen that all the distributions with high p-values have less than 500 events greater than \xmin.  This is because the \xmin\ was so high that it excluded the majority of flares from the fitting process.  This implies that a power law is not appropriate for describing the distribution.

Although  the \lyra\ flare distributions are clearly not power laws, they are typically much steeper than those found using \xrs.  For example, the best fit power law exponent for the distribution in Figure~\ref{fig:lyra_hists}a was found to be $\sim$2.3$\pm$0.1.  For comparison, the exponents for almost all the \xrs\ flare lists derived with both the \goes\ event list and LYRAFF algorithms were less than 2.  In order to determine the cause of this discrepancy, we split the \lyra\ flares into two subdistributions: those observed by both \lyra\ and \xrs\ and those observed by \lyra\ only.  For this exercise we used the \lyra\ flare list derived with the standard LYRAFF algorithm (start threshold=1\%, first fall threshold=20\%, end threshold=50\%) and the standard \goes\ event list (start threshold=40\%, end threshold=50\%).  We found that out of the 5253 flares observed by \lyra\ between 2010 and 2014 inclusive (excluding eclipse seasons), 2276 were also in the \goes\ event list, while 2977 were not.

Next, we verified that the flares not in the \goes\ event list were not bad detections caused by data artefacts or deficiencies in the LYRAFF algorithm.  We randomly selected 100 of the \lyra\ flares   not found in the \goes\ event list and examined their lightcurves.  By visual inspection we determined that 91\% were good detections, while only 9\% appeared not to be a flare.  This confirms that the majority of the LYRAFF detections are real and that there is a population of flares visible to \lyra\ but not detectable by \xrs\ and the \goes\ event list.

Figure~\ref{fig:lyra_hists}b shows histograms of the two \lyra\ subdistributions discussed above.  Flares detected in \lyra\ and \xrs\ observations are represented by the solid line while flares only detected with \lyra\ are represented by the dashed line.  Two things are clear from this plot.  Firstly, the \lyra-only flares tend to be small, which may be a clue as to why they are not detected with \xrs.  Since less intense flares tend to be cooler \citep{ryan2012}, the \lyra-only flares may be too cool to exhibit significant thermal emission in the \xrs\ channels.  Secondly, the \lyra-only flares have a much steeper distribution.  If this were physical it could imply that the flare distribution actually steepens as flares become smaller and would have interesting consequences for the coronal heating problem.  However, another possible explanation is that \lyra's degradation has not been properly corrected and is skewing the results.  

Although \lyra's zirconium channel has experienced less degradation than the other channels, its degradation is still significant.  It is known that this degradation is more severe in the EUV part of the spectral band (6--20\,nm) than the soft X-ray (SXR; $<$2\,nm), which is more sensitive to flare variability.  In order to avoid artificially enlarging the flare variability in the signal -- assumed to be dominated by the SXR contribution -- an additive degradation correction is included in the publicly available data.  However, if the flare variability also has a significant contribution from the EUV part of the passband, then the additive correction would cause flare peak irradiances to be increasingly underestimated as degradation increases.  This would cause the  flare distribution to become steeper over time.  To check this, we plotted histograms in Figure~\ref{fig:lyra_hists}c of the \lyra\ flares not found in the \goes\ event list from 2010 (solid histogram) when the degradation was not too bad and after 2010 (dashed histogram) when the degradation had become more significant.  It can clearly be seen that the 2010 distribution is shallower than that for flares after 2010.  This is consistent with the degradation explanation.  Additionally, it is expected that the difference in slopes would be smaller if all flares observed by \lyra\ 
were included because the flares large enough to also be detected in \xrs\ would have a higher contribution from SXR and the additive degradation correction would affect them less.  This is exactly what we found.  Therefore, we conclude that the steeper slope seen in \lyra\ flare distribution is neither physical nor due to deficiencies in the LYRAFF algorithm.  It is most likely due to insufficient degradation correction of \lyra\ irradiances.  This means that although a combination of \lyra\ and LYRAFF can be used to detect the occurrence of flares, absolute \lyra\ irradiances are not reliable and should not be used for flare statistics or energetics unless the degradation is adequately corrected.

\section{Conclusions}
\label{sec:conc}
In this paper we examined the effect of arbitrary thresholds in the \goes\ event list and LYRAFF flare detection algorithms when applied to \xrs\ and \proba/\lyra\ observations.  We found that that the power law index of the \xrs-derived flare peak flux size distribution depends on the choice of start threshold, but not on those of the first fall and end thresholds.  This was found to be true when both the traditional \goes\ event list and \goes-adapted LYRAFF algorithms were used.  The power law slope value ranged from $\sim$1.92--2.02 depending on whether start thresholds of 90\%--5\% were used.  This highlights the importance of understanding the biases introduced by flare definitions before drawing conclusions on the nature of the derived flare distribution.  This is particularly true if trying to address coronal heating by nanoflares as changing the flare start threshold can cause the flare distribution to have a power law slope either above or below the critical threshold of 2.

We also found that the \lyra\ flare peak irradiance size distribution is artificially steep and not well modelled by a power law.  This was shown to be consistent with an insufficient degradation correction.  Therefore, although \lyra\ can be used to detect flares, its absolute irradiances should not be used for statistics or energetics unless the degradation is adequately corrected.  However, this does not affect \lyra's ability to chart relative irradiance changes over short timescales, for example within flares.  

We consider the discussion in the two previous paragraphs to be the main conclusions of this paper.  However, in the process of this analysis we came to a number of secondary conclusions.  Firstly, we found that that the LYRAFF algorithm does a better job  of separating temporarily overlapping flares than the \goes\ event list.  This leads to a slightly higher proportion of smaller flares being detected.  This may cause the power law slope of the flare peak flux size distribution to steepen.  However, any such steepening observed as part of this study was found to be within the range of statistical uncertainty.  Despite LYRAFF's improved ability to detect temporally overlapping flares, it is still limited.  If a greater number of  such flares could be detected and separated, it is expected that the power law slope of the flare distribution would steepen further.

Secondly, we showed that -- independent of the choice of threshold -- the \goes\ flare peak flux size distribution is visually approximate to but is not strictly a power law.  However, we have not been able to adequately determine the reason for this.  We showed that by assuming the true flare distribution is a power law with an exponent of $\sim$2, it is plausible that the observed distribution's deviation from a power law may be owing to flares being missed by the detection algorithm.  However, if the power law exponent is much steeper than 2, this becomes increasingly unlikely.  The issues with flare detection, solar background, energy partition, and instrumental response suggest that conclusively answering this question with an instrument as basic as \xrs\ may be impossible.

In order to improve our ability to examine the flare distribution, it is necessary to use spatial information.  This will allow flares to be identified when they overlap temporally but not spatially.  It will also improve our ability to accurately subtract the background flux.  Both of these steps will improve our statistics.  The flare distributions from multiple individual active regions should also be studied in this way to determine whether the global flare distribution is a single power law or a convolution of numerous heavy-tailed non-power law distributions.  Studies that more comprehensively chart the energy partition of flares and eruptions must be undertaken to better determine what biases we introduce by using measurements from a single passband as a proxy for the flare energy.  Finally, algorithms that can better identify the event duration will allow us to examine fluences rather than just peak fluxes, which are likely to be a better proxy of the total flare energy.

%
\begin{acknowledgements}
The authors would like to thank Ingolf Dammasch for his helpful discussions.

D.\ Ryan wishes to thank the Solar-Terrestrial Centre of Excellence and the SIDC Data Exploitation project for their financial support.

M.\ Dominique's work has been funded by the Interuniversity Attraction Poles Programme initiated by the Belgian Science Policy Office (IAP P7/08 CHARM).

Support for D.\ Seaton was provided by PRODEX grant No. 4000103240 managed by the European Space Agency in collaboration with the Belgian Federal Science Policy Office (BELSPO) in support of the PROBA2/SWAP mission and by the European Union's Seventh Framework Programme for Research, Technological Development and Demonstration under grant agreement No. 284461 (Project eHeroes, \url{www.eheroes.eu}).

A.\ White has been supported by the the STATICA project, funded by the Principal Investigator programme of Science Foundation Ireland, contract number 08/IN.1/I1879.

\lyra\ is a project of the Centre Spatial de Liege, the Physikalisch-Meteorologisches Observatorium Davos and the Royal Observatory of Belgium funded by the Belgian Federal Science Policy Office (BELSPO) and by the Swiss Bundesamt f\"{u}r Bildung und Wissenschaft.

This research has made use of SunPy, an open-source and free community-developed solar data analysis package written in Python \citep{sunpy2015}.
\end{acknowledgements}

%
%
\bibliographystyle{aa}
\bibliography{Ryanetal_2016.bib}  

%
\begin{appendix}
\section{LYRA Artefacts Removed By LYRAFF}
\label{app:lyra_artifacts}
Table~\ref{tab:lyra_artifacts} shows the different artefacts removed from \lyra\ observations before applying the LYRAFF algorithm.  These artefacts are defined in the \proba\ time annotation files.  For more information see \url{http://proba2.oma.be/data/TARDIS}.

\begin{table*}
\begin{center}
\caption{Standard artefacts removed from \lyra\ observations before applying LYRAFF.}
\label{tab:lyra_artifacts}
\begin{tabular}{ll}     
\hline
Artefact                                &Description \\
\hline
ASIC reload                     &Reload of the application apecific integrated circuit \\
&\\
Calibration                     &Calibration campaign   \\
&\\
Glitch                          &Undefined glitch.      \\
&\\
LAR                                     &\specialcell{Large angle rotation.  90$^o$ rotation of satellite lasting a\\few minutes. Cause spikes and dips in the signal.} \\
&\\
Moon in LYRA                    &\specialcell{Moon in \lyra\ field of view which can cause\\non-solar variations in signal.}       \\
&\\
Offpoint                                &\specialcell{\proba\ offpointed from disk centre potentially altering\\signal artificially.}    \\
&\\
Operational Anomaly             &Unexplained anomaly.   \\
&\\
Recovery                                &\specialcell{Time taken for \lyra\ detectors to get back to full efficiency\\after not being exposed to sunlight. Can take many hours.}           \\
&\\
SAA                                     &\specialcell{Transition of \proba\ through the South Atlantic anomaly\\causing increased noise in signal.} \\
&\\
UV occultation                  &\specialcell{Solar irradiance attenuated by Earth's atmosphere.\\Defined as when \proba's tangential altitude is $\leq$350~km.}  \\
&\\
Visible occultation             &Solar irradiance is blocked by the disk of the Earth.   \\
\hline
\end{tabular}
\end{center}
\end{table*}

\section{Calculating $p$-values}
\label{app:p}
The method of calculating $p$-values used in this study is outlined in Clauset, Shalizi, and Newman (2009). First, we used the methods described in Section 3.3 to find the best fit power law to our empirical data set, giving us an exponent $\hat{\alpha}$, a lower limit $x_{\min}$, and a KS statistic $KS_{emp}$. The total number of flares in the data set is $n$, while the number greater than $x_{\min}$ is $n_{tail}$. Next, we created an ensemble of synthetic data sets. Data points above $x_{\min}$ were randomly generated from a power law with exponent $\hat{\alpha}$, while data points below $x_{\min}$ were randomly sampled with replacement from the corresponding subset of the empirical data. Each synthetic data point had a probability of $n_{tail}/n$ of being in the regime above $x_{\min}$ and $1 - n_{tail}/n$ of being in the regime below this value.  Thus, the synthetic data sets  should mimic the empirical data set well across its entire range if the data above the threshold value of $x_{\min}$ follows a power law. 

The MLE method outlined in Section 3.3 was then used to fit each synthetic data set with its own power law model, and its corresponding KS statistic of the fit, $KS_{synth}$ was calculated. Finally, the $p$-value was given by $p = n_{synth > emp}/n_{synth}$, where $n_{synth>emp}$ was the number of synthetic data sets with a $KS_{synth} > KS_{emp}$ and $n_{synth}$ was the total number of synthetic data sets. In this study we created 1000 synthetic data sets for each empirical one  so that  our $p$-value would be reliable to two decimal places \citep{clau2009}.

\end{appendix}
\end{document}